\documentclass[sigplan,10pt]{acmart}

\setcopyright{none}

\renewcommand\footnotetextcopyrightpermission[1]{}

\AtBeginDocument{%
  \providecommand\BibTeX{{%
    \normalfont B\kern-0.5em{\scshape i\kern-0.25em b}\kern-0.8em\TeX}}}

\usepackage{times}
\usepackage{titlesec}
\usepackage{outlines}
\usepackage{enumitem}
\usepackage{amsthm}
\usepackage{nameref}
\setenumerate[1]{label=\Roman*.}
\setenumerate[2]{label=\Alph*.}
\setenumerate[3]{label=\roman*.}
\setenumerate[4]{label=\alph*.}
\usepackage{dblfloatfix}
\usepackage{flushend}

\usepackage[T1]{fontenc}
\usepackage[scaled=0.85]{beramono}

\usepackage{hyperref}

\usepackage{multirow}
\usepackage{graphicx}

\usepackage{amsmath}

\usepackage{xcolor}
\usepackage{refcount}

\usepackage[noend]{algpseudocode}
\usepackage{algorithm}
\usepackage{algorithmicx}

\usepackage{microtype}
\usepackage{times,epsf,epsfig,textcomp,amsmath}
\usepackage{wrapfig}
\usepackage{comment}
\usepackage{array}
\usepackage{colortbl}
\usepackage{url}
\usepackage{color, soul}

\usepackage{graphics}

\usepackage[labelfont=bf,font={small}]{caption}
\usepackage{subcaption}
\usepackage{float}

\usepackage{multirow}
\usepackage{capt-of}
\usepackage{enumitem}
\usepackage{siunitx}
\usepackage{listings, listings-rust}
\usepackage{stmaryrd}

\usepackage{bm}
\usepackage[mathscr]{euscript}
\usepackage{xspace}

\usepackage{helvet}
\usepackage{tgheros}
\usepackage[section]{placeins}
\newcolumntype{L}[1]{>{\raggedright\let\newline\\\arraybackslash\hspace{0pt}}m{#1}}
\newcolumntype{C}[1]{>{\centering\let\newline\\\arraybackslash\hspace{0pt}}m{#1}}
\newcolumntype{R}[1]{>{\raggedleft\let\newline\\\arraybackslash\hspace{0pt}}m{#1}}

\newcommand\Snospace[1]{}
\def\myS~{\S{}}

\def\myL~{L{}}

\definecolor{lightgray}{gray}{0.6}
\definecolor{lightblue}{rgb}{0.9,0.9,1}
\definecolor{aqua}{rgb}{0.0, 1.0, 1.0}

\newcommand{\Paragraph}[1]{\vskip 6pt\noindent\textbf{#1. }}
\renewcommand{\paragraph}{\Paragraph}
\newcommand{\itemno}[1]{(\textit{#1})}

\makeatletter
\lst@CCPutMacro
    \lst@ProcessOther {"2A}{%
      \lst@ttfamily 
         {\raisebox{1.5pt}{*}}
         \textasteriskcentered}
    \@empty\z@\@empty
\makeatother

\lstdefinestyle{mystyle}{
    basicstyle=\fontsize{6.5}{7.2}\selectfont\ttfamily,
    commentstyle=\color[rgb]{0.13,0.55,0.13}, 
    keywordstyle=\bfseries\color{blue},        
    keywordstyle=[2]\color[rgb]{0,   0.5, 0.5  },   
    keywordstyle=[3]\color[rgb]{0.5, 0,   0    },   
    keywordstyle=[4]\color[rgb]{0,   0.5, 0    },   
    keywordstyle=[5]\color[rgb]{0,   0,   0.75 },   
    numberstyle=\tiny\color{darkgray},
    stringstyle=\color{red},
    breakatwhitespace=false,         
    breaklines=true,                 
    captionpos=b,                    
    keepspaces=true,                 
    numbers=left,                    
    numbersep=-2pt,                  
    showspaces=false,                
    showstringspaces=false,
    showtabs=false,                  
    tabsize=2,
    upquote=true,
    escapechar=\$,
}

\usepackage{tikz}

\newcommand\whitecircle[1]{\small \raisebox{0.1em}{\textcircled{\footnotesize \bfseries \raisebox{-0.7pt}{#1}}}}

\newcommand\couldremove[1]{{\color{lightgray} #1}}
\newcommand{\remove}[1]{}

\newcommand{\code}[1]{{\small \texttt{#1}}}
\newcommand{\program}[1]{\textsf{\small{#1}}}

\renewcommand\couldremove[1]{}  

\newcommand{\pt}[1]{{\textbf{\textsf{\scriptsize{#1}}}}}

\newcommand{\name}{Theseus\xspace}

\newcommand{\cert}{strength of guarantee\xspace}

\newcommand{\ICS}{IRS\xspace}
\newcommand{\capi}{representation\xspace}
\newcommand{\Capi}{Representation\xspace}
\newcommand{\capis}{representations\xspace}
\newcommand{\Capis}{Representations\xspace}

\newcommand{\lem}{\textsf{\small Lemma}\xspace}
\newcommand{\lems}{\textsf{\small Lemmas}\xspace}

\usepackage{amsthm}

\newtheoremstyle{lemma}
{2ex}
{2ex}
{}
{9pt}
{\bfseries}
{. }
{}
{}

\theoremstyle{lemma}

\newcommand{\RepresentationCreator}{RepCreator\xspace}

\begin{document}



\title{{\LARGE Combining Type Checking and Formal Verification for Lightweight OS Correctness}}

\author{Ramla Ijaz}
\email{ramla.ijaz@yale.edu}
\affiliation{%
  \institution{Yale University}
  \country{}
}

\author{Kevin Boos}
\email{kevinaboos@gmail.com}
\affiliation{%
  \institution{Theseus Systems}
  \country{
  }
}

\author{Lin Zhong}
\email{lin.zhong@yale.edu}
\affiliation{%
  \institution{Yale University}
  \country{}
}

\begin{abstract}
This paper reports our experience of
providing lightweight correctness guarantees to an open-source Rust OS, Theseus. First, we report new developments in intralingual design that leverage Rust's type system to enforce additional invariants at compile time, trusting the Rust compiler. Second, we develop a hybrid approach that combines formal verification, type checking, and informal reasoning, showing how the type system can assist in increasing the scope of formally verified invariants.
By slightly lessening the strength of correctness guarantees, this hybrid approach \emph{substantially} reduces the proof effort. 
We share our experience in applying this approach to the memory subsystem and the 10 Gb Ethernet driver of Theseus, demonstrate its utility, and quantify its reduced proof effort.

\end{abstract}

\maketitle
\pagestyle{plain} 
\vspace{3mm}
\section{Introduction}
\label{sec:intro}
Correctness is a desirable yet challenging property to achieve for systems software such as an operating system (OS). A key technology for correctness is formal verification. 
In recent years, various formal verification approaches have emerged that make different trade-offs between expressiveness and proof effort, i.e., what can be proven vs. how difficult it is to generate those proofs (\S\ref{sec:background}).

This paper presents our experience exploring new ways to ensure OS correctness. Toward achieving high expressiveness with low proof effort, we relax the \emph{strength of correctness guarantees}. 
We observe that while full formal verification is desirable, the type system of the implementation language, i.e., Rust, combined with informal reasoning can also be used to provide weaker yet distinctly useful guarantees.

In \S\ref{sec:intralingual}, we further develop the idea of \emph{intralingual design} introduced by Theseus OS~\cite{boos2020osdi}, expanding its reach with new techniques. Intralingual design uses language-level features to enforce invariants via the compiler.
We present the idea of a \emph{representation}, a linear type instance that is the sole means of accessing a system resource, and show how to use language features to shift resource management responsibilities into the compiler.
We then show how to use linear type instances as a proof of work, which can enforce correct ordering for operations.
Lastly, we present how to write an intralingual Hardware Abstraction Layer (HAL) that statically prevents bugs at the hardware programming interface.

We analyze the limits of intralingual design, showing that the invariants presented in~\cite{boos2020osdi} were based on an incomplete foundation:
while Rust's ownership model guarantees that an instance of a linear type has a single owner, it cannot guarantee the absence of overlap between the {\em values} of two separate instances of the same linear type.
This shortcoming led to an insidious bug in the original memory subsystem of Theseus~\cite{boos2020osdi}, for which we report and contribute a solution.

Motivated by the limitation of intralingual design, we advocate a hybrid approach that combines intralingual design, formal verification, and informal reasoning to achieve \emph{lightweight} correctness for OSes (\autoref{sec:hybrid}).
We consider it lightweight not only because it requires much less effort compared to conventional formal verification, but also because its \cert is weaker due to usage of informal reasoning and implicit trust of the implementation language. 
We aim to maximize use of the type system through intralingual design, as it is a low-effort way to realize stronger correctness guarantees. 
To that end, we introduce the idea of \emph{intralingual specifications} which allow the compiler to check type-related correctness properties.
Then, when proving an invariant, we only apply formal verification where said properties cannot be upheld by the type system, where the increased guarantee justifies the proof effort. Since formal verification is expensive, we present three rules that help to increase the reach of formally-verified invariants using the type system.

In \autoref{sec:impl}, we report our experience in applying this approach to revise the memory subsystem of Theseus and to implement a driver for the Intel 82599 NIC.
In the memory subsystem, we create representations of the \code{Pages} and \code{Frames} types, which are fundamental to memory management in \name.
Using our hybrid approach, we verify functions that create these representations and reason about how this leads to stronger guarantees of the original \name invariants.
In the process, we eliminate an insidious class of bugs. 
In the network driver, we show that extending intralingual design can avoid common driver bugs and that our hybrid approach can uphold core invariants with less proof effort than end-to-end verified drivers.

We evaluate the ``lightweightedness'' of our approach and quantify any performance overhead in \S\ref{sec:evaluation}. We show that the hybrid approach has low development burden: the proof-to-implementation ratio is 1:117 for the memory subsystem, and 1:8.3 for the \program{ixgbe} driver.
We find that our hybrid approach has negligible performance overhead:
our \program{ixgbe} driver performs similarly to other research drivers with correctness guarantees, with a maximum throughput only 5\% lower than the DPDK \program{ixgbe} driver.
We also use the intralingual HAL of our \program{ixgbe} driver to find previously unreported bugs in \program{ixgbe} drivers from other Rust-based OSes such as ixy~\cite{emmerich2018ixy}, Redox~\cite{redox}, and RedLeaf~\cite{narayanan2020redleaf}. 

In summary, this paper makes three contributions: 
\begin{itemize}[topsep=0.3em,itemsep=0.02em,leftmargin=1em]
    \item A new set of techniques to expand the scope and extend the reach of intralingual design.
    \item  A hybrid proof approach that combines type checking (as used by intralingual design), formal verification, and informal reasoning to improve OS correctness.
    \item Our experience applying this low-effort approach to increase the proven correctness of a real system, namely the memory subsystem and \program{ixgbe} driver of \name. 
\end{itemize}

\section{Background and Related Work}
\label{sec:background}

\paragraph{Rust's Type System}
A language type system is a lightweight formal method for encoding program behaviors.
The Rust programming language employs a linear (technically, {\em affine}) type system~\cite{pierce2004advanced} 
in which a variable can only be \textit{used at most once}.
This prevents aliasing by restricting each instance to one owner, represented by a variable binding that "owns" the underlying memory.
An instance of a linear type cannot be duplicated via copying or cloning;
rather, ownership of that instance can only be assigned (moved) to another variable binding, preventing the prior owner from using it again.

Rust's linear type-based ownership model allows memory usage and aliasing to be statically determinable in most cases.
The compiler can track the lifetime of an instance and insert code, i.e., \textit{a drop handler}, to reclaim it when its owner's scope ends. 
As a result, Rust programs are both memory and concurrency safe without underlying runtime or garbage collection, achieving performance close to other systems programming languages like C and C++. Not surprisingly, it has become popular in systems programming in recent years, including implementing operating systems~\cite{levy2017sosp,boos2020osdi,narayanan2020redleaf}.

In addition to ownership, Rust allows instances to be temporarily "borrowed" by another variable binding (reference) without transferring ownership.
Rust enforces aliasing XOR mutability, wherein there can only exist one mutable reference or multiple immutable references to an owned instance at a given time, but not both.

\paragraph{Rust for Correctness}
Many have leveraged Rust's type system to ensure correctness beyond safety. We next briefly review these ideas before developing them further in~\S\ref{sec:intralingual}.

\textit{Linear Types for Pairwise Operations}:~~
Many operations must always occur in pairs, e.g., memory allocation/deallocation, lock acquisition/release, and reference count incre\-ment/decrement.  
Mismatchings of such pairwise operations are common in the Linux kernel ~\cite{xu2015asplos,mao2016asplos,liu2022usenixsecurity}.
This problem can be solved by using linear types, placing the first operation in the constructor and the second in the destructor.
Rust itself follows this design pattern for heap-allocated data structures (\code{Vec<T>}~\cite{rust_vec}), locks (\code{MutexGuard<T>}~\cite{rust_mutex}), and reference-counted pointers (\code{Arc<T>}~\cite{rust_arc}).

\textit{Linear Types as Unforgeable Capabilities}:~~
In Rust, an instance of a linear type is a \emph{unique} and \emph{unforgeable} capability as long as it does not implement the \code{Clone} trait, meaning it cannot be duplicated~\cite{narayanan2020redleaf, kulkarni2018splinter}.
The ownership of such an instance automatically confers the right to use it without the need for runtime checks of its authenticity~\cite{narayanan2020redleaf, fahndrich2002adoption-pldi}.
Since a capability is of a linear type, it has a single owner, and Rust's built-in ownership rules can prevent data races and automatically insert destructors.
In \S\ref{sec:intralingual_ics}, we take inspiration from this idea by representing OS resources with linear-type instances.

\textit{Linear Types for Statically-Enforced State Machines}:~~
A linear type system can prevent incorrect state machine transitions by implementing the state machine using behavioral type techniques, e.g., typestates and session types.
When combined with linear types, a typestate protocol can be statically validated~\cite{fahndrich2002adoption-pldi}, imposing no runtime overhead. 

\paragraph{Formal Verification}
Formal verification is expensive; developers often limit expressiveness, i.e., what can be proven, in order to control the cost.
A formally-verified system consists of three parts: implementation, specification, and proof. 
Various formal verification approaches have emerged that make different trade-offs between expressiveness and proof effort, i.e. what can be proven vs. how difficult it is to generate those proofs.
\emph{Interactive theorem proving} is the most expressive, as it can reason about higher-order logic but suffers from the largest proof effort, measured by the proof-to-implementation ratio, 13: 1 for CertiKOS~\cite{gu2016certikos}. 
By employing SMT solvers to find proofs, later works were able to substantially lower the proof effort at the cost of limiting proof requirements to first-order logic.
However, even so-called \emph{push-button} approaches~\cite{sigurbjarnarson2018nickel-osdi,sigurbjarnarson2016osdi,nelson2017hyperkernel-sosp,nelson2019sosp, zaostrovnykh2019sosp} still suffer from significant proof effort, even for very small system software and for proving limited invariants, mainly restricted to the decidable portion of first-order logic.
For example, Serval~\cite{nelson2019sosp}, implemented in 2K LoC, requires an additional 3.1K LoC for its specification and verifier tools.
Our hybrid approach aims to further lower this proof effort by exploiting the type system of the implementation language and informal reasoning, at the cost of lowered \cert.

\paragraph{Other Related Work}
Our work is complementary to the literature that also exploits linear types in systems software for other purposes. 
The Singularity project~\cite{hunt2007singularity} popularized linear types for OS design and used them for zero-copy sharing of heap memory across software-isolated domains.
Recent works have used Rust's ownership model for features such as lightweight fault isolation \cite{narayanan2020redleaf}, zero-copy communication \cite{panda2016netbricks-osdi}, compiler-checked session types \cite{jespersen2015session}, decentralized resource management \cite{boos2020osdi}, and static information flow control and automatic program state manipulation~\cite{balasubramanian2017hotos}. 

Related to our hybrid approach, Yang and Hawblitzel  creatively combined formal verification and a safe implementation language by dividing the OS into a lower core \textit{Nucleus} and a higher kernel in building the Verve OS~\cite{yang2010safe}. 
They applied formal verification to the Nucleus, implemented in assembly, for safety and correctness, while relying on the implementation language (C\#) for the kernel's safety. 
Our hybrid approach does not mandate a strict division between verified and unverified OS portions.
Instead, we use a combination of proof techniques in whichever subsystem we aim to prove an invariant about, pairing each correctness property with the proof technique best suited to it.
Our work designs and implements the OS so that the Rust type system can provide guarantees that go beyond safety, in collaboration with formal verification.

Related to our application of intralingual design to Theseus's \program{ixgbe} driver, Vigor~\cite{zaostrovnykh2019sosp} found errors in the DPDK ixgbe driver by symbolically executing it against an 82599 hardware model, using assertions to catch bit-level errors. 
TinyNF~\cite{pirelli2020tinynf-osdi} introduced a simplified driver model with fewer code paths, enabling faster verification of network functions that run on top of it. 
Ironclad~\cite{hawblitzel2014ironclad} verified a 1 Gbps Ethernet driver using the Dafny verification language.
Unlike these approaches, our method proves driver correctness using a combination of the Rust type system and verification, which reduces both specification and proof effort.

\section{Intralingual Design}
\label{sec:intralingual}

Intralingual design \cite{boos2020osdi} aims to maximize the compiler's role in enforcing correctness by leveraging programming language features, namely type systems, to more precisely convey system requirements to the compiler. 
In other words, it \emph{encodes the requirements} into the implementation itself, such that they can be enforced by the compiler. 
Many requirements cannot be so encoded because the type system is not expressive enough to convey them. We categorize these requirements as \emph{extralingual}. 
Our objective herein is to build upon recent works on Rust for Correctness (\autoref{sec:background}) and introduce a \textit{systematic methodology} for incorporating linear types and other type-based techniques into low-level system design.

We leverage linear types in three ways.
(\textit{i}) First, we use linear types to create an \textit{exclusive} \capi for a resource, both physical and virtual (\S\ref{sec:intralingual_ics}).
(\textit{ii}) Second, we use linear types as a \textit{proof of work} by using distinct types for function return values and preventing said types from being otherwise instantiated via any other code path (\S\ref{sec:intralingual_proof_of_work}).
(\textit{iii}) Third, we employ type-system techniques to enforce datasheet-compliant communication with the hardware (\S\ref{sec:intralingual_hal}). 

\begin{figure}[t]
    \begin{center}
    \begin{lstlisting}%
        [language=Rust, 
        xleftmargin=.03\textwidth,
        belowcaptionskip=-5.5mm,
        % float, 
        %frame=single, 
        caption={An example of using an \ICS to manage virtual memory. A \code{Pages} instance is a \capi of a range of pages, which can be in one of four states. When a \code{Pages<Mapped>} instance is dropped, it is eventually returned to the \code{Free} state and stored in the list of free pages. The code has been simplified for brevity/readability.},
        label={lst:intralingual_design}]%
    %
    
    // Pages is a representation of a range of virtual pages.
    struct Pages<S: State> { $\label{line:pages}$
       range: RangeInclusive<usize>
    }
    // The possible states a Pages instance can be in.
    enum State { $\label{line:mem_states}$
       Free,
       Allocated,
       Mapped,
       Unmapped
    }
    // Only Pages in the Mapped state can access memory.
    impl Pages<Mapped> {
       pub fn write(&mut self, data: [u8]); $\label{line:pages_write}$
       pub fn read(&self) -> &[u8]; $\label{line:pages_read}$
       fn unmap(self) -> Pages<Unmapped>; $\label{line:pages_into_unmapped}$
    }

    impl<S: State> Drop for Pages<S> {
       fn drop(&mut self) { $\label{line:pages_drop}$
          match S {
             State::Free => { $\label{line:pages_drop_free}$
                // Re-take ownership of the pages by replacing it 
                // with an empty range; return it to page allocator.
                let pages = replace(&mut self, Pages::empty());
                free_pages_list.insert(pages);
             } 
             State::Mapped => { $\label{line:pages_drop_mapped}$
                // PTE(s) have been cleared, so we transition 
                // the Pages to the Unmapped state.
                let pages = replace(&mut self, Pages::empty());
                pages.unmap(); // Drop the returned Pages<Unmapped>
            } ...
          }
       }
    }
\end{lstlisting}%
\end{center}
\end{figure}

\subsection{Intralingual \Capi System}
\label{sec:intralingual_ics}
                
We present an Intralingual \Capi System (\ICS) that shifts some of the responsibility of managing system resources from the OS to the compiler.
Typically, an OS creates software objects to represent physical or abstract resources, e.g., Linux uses \code{struct page} objects to represent physical memory frames. 
The \ICS combines the representation of a resource with the authority to use it, through linear types:
the ownership of the linear-type instance denotes the sole authority to use the resource.
We call this instance a {\em \capi } of the resource. 
\Capis in an \ICS are checked by the compiler, which ensures that \itemno{i} there is only ever one mutable reference to the \capi at a time, and
\itemno{ii} access to the \capi is governed by the rules conveyed via the type system. 
We note that Rust already use linear type instances as a limited form of \capis for memory objects, but \ICS extends this to apply \capi types to arbitrary system resources beyond just memory.
We realize the following key features of an \ICS:

\paragraph{Changing Access Rights via Typestates}
The access rights of a \capi are defined by publicly visible methods of its type. Each typestate represents a distinct set of access rights, which change with state transitions. 
A state transition method takes a \capi as input, \textit{consumes} it, and changes its state.
For example, in \autoref{lst:intralingual_design}, a \code{Pages} instance is a \capi that can be in one of four states: \code{Free}, \code{Allocated}, \code{Mapped}, or \code{Unmapped}~(\autoref{line:mem_states});
its methods transition the \capi between these states, e.g., \code{unmap()} in \autoref{line:pages_into_unmapped}.
The compiler can enforce that the \capi is accessed according to the restrictions of its current state.
For example, in the \code{Free}, \code{Allocated}, and \code{Unmapped} states, page table entries (PTEs) are not set up for the given pages, so the \code{Pages} \capi cannot be used to access the underlying memory range.
This is statically enforced by implementing \code{read()} and \code{write()}  \textit{only} for \code{Pages} in the \code{Mapped} typestate~(\autoref{line:pages_write}).

\paragraph{Delegation via Ownership Transfer, Sharing, Borrowing}
A \capi is a singleton with either one exclusive owner or multiple owners that can only mutably access it through a mutual exclusion mechanism, upholding Rust's aliasing XOR mutability (\S\ref{sec:background}).
An owner can conveniently delegate authority by granting access to the \capi in one of three ways:
\itemno{i} transferring ownership to a new owner who gains exclusive access,
\itemno{ii} sharing ownership via a reference-counted smart pointer so multiple parties can co-own the \capi, or 
\itemno{iii} temporary (scoped) lending to a borrower that can access the \capi through a reference.

\paragraph{Returning \Capis via Automatic Destructors}
We can shift the complex responsibility of realizing correctly-ordered cleanup sequences from the programmer to the compiler by placing all cleanup code in a linear type's destructor (a Rust \code{Drop} handler).
This is important for \capis that represent physical resources (e.g., physical frames) that should never be destroyed: these \capis must be returned to the OS for future use. 
With typestates, a \capi can have multiple drop handlers, one for each state; each state's drop handler undoes any changes made when entering that state, reverting the \capi to its previous state.
The drop handler for the initial state finally returns the instance back to the OS for storage.
For example, in \autoref{lst:intralingual_design}, the drop handler for \code{Pages} in the \code{Mapped} state removes the PTEs and converts it to the \code{Unmapped} state (\autoref{line:pages_drop_mapped}).
Then, each predecessor state's drop handler is iteratively invoked until the \code{Pages} instance returns to the \code{Free} state, upon which the \code{Pages<Free>} instance is returned to a redblack-tree of free page chunks maintained by the page allocator (\autoref{line:pages_drop_free}).

\vspace{0.5em}\noindent \textit{\Capi vs. Capability}:~~
The notion of a \capi may remind the readers of that of a capability.
Like a capability, a \capi is also \textit{unforgeable} and \textit{delegable}. Unlike a capability, a \capi is \textit{unique} in that no two \capis exist in the system for the same resource. This precludes \emph{derivation} in which multiple copies of a capability, with varying access rights, exist at the same time.
Importantly, in-built language features do not provide all features of a capability system as reported by~\cite{von1999jkernel, RobustComposition}. 
Instead the OS must use language-level mechanisms to implement these features, such as revocation via a level of indirection.


\subsection{Linear Types as Proof of Work}
\label{sec:intralingual_proof_of_work}
The other main way we use linear types is to indicate that a certain function has been executed, by returning a dedicated type instance from the function.
In this manner, a linear type instance no longer represents a \textit{spatial} resource, but rather a proof of a \textit{temporal} action having occurred.
A linear type used for this purpose is simply a type that can only be instantiated by a single function that first performs the required ``work.''
To prevent instances of this type from being created anywhere, we ensure it is composed of a \emph{private} inner type that is inaccessible outside of the type's module.

Combining \emph{linear types as a proof of work} with strongly-typed function interfaces can statically enforce an order between ``stages'' of operations.
That is, we can create a chain of functions where one function creates and returns an instance of a linear type to be consumed by the next function, effectively requiring each instance of a linear type to be used in the order it is created for progress to be made.
In fact, this pattern of instantiation and then consumption lies at the core of many ways in which we use linear types (\autoref{sec:background}).
The difference here is in what the type instance represents.

\begin{figure}[t]
    \begin{center}
    \begin{lstlisting}%
        [language=Rust, 
        xleftmargin=.03\textwidth,
        belowcaptionskip=-5.5mm,
        % float, 
        %frame=single, 
        caption={A portion of the Intel 10 GbE driver intralingual HAL. The padding between the registers which forces them to lie at their correct offset in the MMIO region is ommitted for space.},
        label={lst:intralingual_hal}]%
    %

    // register struct which is mapped to the MMIO region
    pub struct IntelIxgbeRegisters {$\label{line:intel_ixgbe_regs}$
        rxctrl:     ReadWrite<u32>, $\label{line:rxctrl}$
        pub gprc:   ReadOnly<u32>, $\label{line:gprc}$
        fctrl:      ReadWrite<u32>, $\label{line:fctrl}$
    }

    // linear types that serve as a proof of work
    pub struct RxCtrlDisabled(()); $\label{line:rxctrldisabled}$
    pub struct FilterCtrlSet(()); $\label{line:fctrlset}$

    // filters that can be enabled
    bitflags! {
        pub struct FilterCtrlFlags: u32 { $\label{line:bitflags}$
            const STORE_BAD_PACKETS             = 1 << 1;
            const MULTICAST_PROMISCUOUS_ENABLE  = 1 << 8;
            const UNICAST_PROMISCUOUS_ENABLE    = 1 << 9;
            const BROADCAST_ACCEPT_MODE         = 1 << 10;
        }
    }

    // register access methods
    impl IntelIxgbeRegisters {
        pub fn rxctrl_rx_disable(&mut self) -> RxCtrlDisabled;
        pub fn fctrl_write( $\label{line:fctrl_write}$
            &mut self, 
            val: FilterCtrlFlags, $\label{line:fctrl_write_flags}$
            rx_disabled: RxCtrlDisabled
        ) -> FilterCtrlSet;
        pub fn rxctrl_rx_enable(&mut self, fctrl_set: FilterCtrlSet)$\label{line:rxctrl_rx_enable}$
    }\end{lstlisting}%
    \end{center}
\end{figure}

\subsection{Intralingual Hardware Abstraction Layers}
\label{sec:intralingual_hal}

An intralingual Hardware Abstraction Layer (HAL) enforces datasheet-provided rules for communicating with a given hardware device at compile-time. 
It is system-independent code that is reusable across drivers written in the same language.
An intralingual HAL uses basic type system features such as \code{structs} with associated methods, type wrappers, visibility modifiers, and \code{enums} to restrict access to MMIO fields and limit what can be written to them.
It also uses linear types as a proof of work to enforce inter-field dependencies. 

The core part of the intralingual HAL is a \code{struct} to represent the layout of memory-mapped I/O (MMIO) registers and other I/O data structures, which can then be overlaid atop a region of memory.
This ensures that every register and bitfield is always accessed in a type-safe manner at its correct offset (and alignment) within the underlying memory region.
Our approach precludes the unsafe pointer arithmetic commonly used to access MMIO registers, which cannot be reasoned about by the compiler.
In \autoref{lst:intralingual_hal}, the \code{IntelIxgbeRegisters} \code{struct} (\autoref{line:intel_ixgbe_regs}) is part of the intralingual HAL for the Intel 82599 NIC. It defines the layout of the memory-mapped registers taken from the datasheet~\cite{82599datasheet}.

For portability, an initial version of the HAL may only use Rust primitive types or types that are also part of the HAL. But a developer can tailor the HAL so that it references system-specific types that carry an invariant, and that invariant serves as a valid pre-condition for a HAL function. For example, the \code{set\_entry} function for a PTE in \name consumes the type \code{AllocatedFrame} rather than a \code{u64} physical address. \code{AllocatedFrame} carries the invariant that there is no existing PTE for that frame, so it helps uphold the requirement that \name only adds PTEs for unmapped frames.

\subsection{Limitations of Intralingual Design}
\label{sec:limit}
Intralingual design, while powerful, is limited due to its reliance on the language's type system. 
Firstly, it cannot reason about unrestricted types: where they originated from or the validity of their values.
This limitation is fundamental to an OS, in which the lowest layers must use built-in unrestricted/ primitive data types like \code{u32} in order to interact with hardware.

Moreover, as the \ICS design uses a linear-type instance to represent an OS resource, \emph{a linear type system itself cannot guarantee uniqueness of the resource represented}.
This goes beyond the intralingual (type-level) uniqueness based on Rust's ownership model: no two variables of the same linear type can own the same value (memory object), 
but there is no guarantee that the resources represented by the values (memory objects) 
do not overlap. 
If there is an overlap, multiple instances of this type can give access to the same resource, i.e., the overlapping parts.
This limitation underlies our discovery of an important bug in \name's memory subsystem, discussed in \autoref{sec:bug}.

Finally, a linear type system is not as expressive as many formal verification techniques due to the limited invariants that can be enforced by the type system.
Generally, it is incapable of proving any algorithmic property, e.g., that a \code{sort()} function actually performs sorting. 

\section{Hybrid Approach for Correctness}
\label{sec:hybrid}

To overcome the limitations of intralingual design and the high proof effort of formal verification, we argue for a hybrid approach that pairs a correctness property with a proof technique.
We maximize the use of the type system because it gives a high \cert with significantly lower proof effort than formal verification. 
With intralingual design, the proof effort is basically the implementation effort.
We use the SMT-based formal verification to prove select, important properties.
Unlike \emph{end-to-end verification} common in most formally verified systems, we employ \emph{selective verification} and rely on the type system to carry forward proven invariants throughout the system.
Manual code inspection is only used in a few simple cases where the type system is limited.

We also embrace informal reasoning, especially prose proofs, to reason about higher-level invariants. 
Prose proofs can ``stitch'' together invariants proven by different proof techniques (and those specified in different languages) without requiring a unified formal specification, significantly lowering the proof effort.
Within a prose proof, we use natural language to express the invariants and justify how they combine to imply a high-level invariant.

Given that a linear-type system itself cannot guarantee uniqueness of the resource represented by an instance, and that such uniqueness is the foundation of an \ICS (\S\ref{sec:intralingual_ics}), the uniqueness of a linear-type instance is an excellent candidate for formal verification. 
To formally verify that a linear-type instance is unique, we only need to formally verify that the type's constructors will never create instances that represent overlapping resources.

\subsection{Rules to Extend the Reach of Verified Invariants}
\label{sec:hybrid-rules}
As formal verification is the main source of proof effort in our hybrid approach, we give three rules to leverage the type system to extend the reach of formally-verified guarantees. 

\textit{(i) Rule of Invariant Preservation:} Once we have used formal verification to prove an invariant for a given linear type instance, the type system ensures that invariant will hold for the instance's lifetime.
No further verification is needed as long as arbitrary changes to the type's fields are disallowed.
A proven invariant for all instances of a type is called a \emph{type invariant}.
Type invariants act as implicit pre-conditions when that type is used as a function argument and post-conditions when it is used in a return value, reducing the lines of specification we have to write.

\textit{(ii) Rule of Composition:} An instance of a composite type is unique if its members are unique. 
Therefore, it is unnecessary to formally verify uniqueness as a type invariant for every \capi in an \ICS.
Instead, by verifying the uniqueness property of select basic types in the system, e.g., \code{Pages} in \autoref{lst:intralingual_design}, we can rely on the type system to propagate
and automatically enforce uniqueness for composite types built from these verified basic types.
For example, a \capi of a device's MMIO registers is composed solely of a \code{Pages<Mapped>} instance; thus, the MMIO \capi is unique because \code{Pages<Mapped>} is unique.

\textit{(iii) Rule of Reuse:} Using Rust generics, we can formally verify that a piece of generic code upholds an invariant, and the compiler automatically extends this verification to all specialized instances.
As an example, after writing and verifying just one implementation of a generic representation creator, we can then use it to create many other verified representations: \code{Pages}, \code{Frames}, and \code{PCIDevice}.
In general, we strive to verify code at lower layers of the system, with more generic/ primitive types to increase the reusability of said formal proof.
Types composed from these primitive types can directly call the verified methods.

\subsection{Intralingual Specifications}
\label{sec:hybrid-intralingual-spec}

While the type system can ensure certain properties, it alone cannot prove general correctness of the system implementation.
Taking inspiration from how specifications used in formal verification {\em can} prove correctness, we introduce a hybrid proof technique called \emph{intralingual specifications}.
We realize these by leveraging Rust macros to specify correctness properties about a given type {\em directly} in the code, such that they are automatically checked at compile time; this increases their \cert with low effort.

The type-based properties that occur most frequently (in our experience) are that:
(\textit{i}) a type must not implement certain traits,
(\textit{ii}) a type is composed of another type,
(\textit{iii}) a function consumes an instance of a type,
(\textit{iv}) a type's inner fields are private.
For example, instances of the \code{Pages} type in \autoref{lst:intralingual_design} must be unique. 
A proof of uniqueness here requires that the type is linear and does not expose its inner fields.
These requirements can be coded by implementing neither the \code{Clone} nor \code{DerefMut} traits for \code{Pages}, and by setting its \code{range} field to private, respectively.
A change in the codebase, e.g., a heedless developer implementing \code{Clone} for \code{Pages}, could inadvertently break the proof of uniqueness. 
Since said changes would not violate Rust type rules, the compiler alone could not catch them.

To check such properties at compile time, we employ Rust macros to write static assertions. These assertions generate code that aborts compilation if the condition is not satisfied.
We have written a tool to collect these assertions into a separate file from the code base to make it easier for a developer to review. 
For example, to prevent implementation of traits for the \code{Pages} type, we can add this line to \autoref{lst:intralingual_design}:\\
\centerline{\small \code{assert\_not\_impl\_any!(Pages: DerefMut, Clone);}}\\    
To ensure that the inner field of \code{Pages} is private, we add this attribute to the top of the \code{Pages} struct:\\
\centerline{\small \code{\#[private\_fields("range")]}}\\ 
We also implement additional attributes,  \code{\#[nomutates]} and \code{\#[nocalls]}, to help prevent unverified functions from breaking invariants proven by verified functions in the same module.
These attributes are placed at the top of a function to blacklist the argument fields that should not be mutated and functions that should not be called.
Standard Rust can only enforce visibility modifiers at a module boundary, but these attributes can enforce visibility restrictions {\em within} a module. Our implementation of these macros is available in the \code{proc-assertions} crate on the Rust crate registry~\cite{procassertions}.

\subsection{Increase in Trusted Code}
Selective verification leads to increased amounts of trusted code. Within a module, we trust that unverified code will not break invariants proven by verified code. For every dependency of verified code, we must write a trusted specification so that the verifier can reason about its behavior. In our work, we trust the Rust core and alloc libraries and have written trusted specification for multiple functions  in the \code{mem}, \code{num}, \code{cmp}, \code{option}, \code{result} and \code{vec} modules.
We also add trusted specification to other kernel crates in Theseus.
In contrast, end-to-end verified systems have much smaller trust code. 

\section{Implementation}
\label{sec:impl}

We applied intralingual design (\autoref{sec:intralingual}) and our hybrid correctness approach (\autoref{sec:hybrid}) to parts of \name OS. For each subsystem we modified, we identified the invariants of interest, implemented an IRS and intralingual HAL. We then used Prusti~\cite{astrauskas2019leveraging}, a Rust-based SMT-verification tool, to formally verify select functions which would help uphold the invariants along with the intralingual code.
Our goal was to maximize reliance on the type system and compiler while minimizing formal verification, either by applying our rules from \autoref{sec:hybrid-rules} or by writing intralingual specifications (\autoref{sec:hybrid-intralingual-spec}).

To address the primary limitation of an IRS, the lack of a uniqueness guarantee, we wrote a verified generic interface that creates representations. We then used this interface to instantiate representations for both \code{Frames} and \code{Pages}, which form the basis of memory management in \name. The uniqueness proof of \code{Pages} and \code{Frames} is essential to upholding the bijective mapping invariant of \name: \emph{each page in the system's virtual address space can only be mapped to one frame of the physical address space, and vice versa.}
This prevents extralingual aliasing, which is necessary to realize memory isolation and safety for \textit{all} system memory, not just the heap and stack. 
In the network driver, we build upon said memory invariants to show how our implementation upholds three correctness properties of a network driver: correct bit-level communication, software resource management, and bookkeeping of hardware state.

\begin{figure}[t]
	\centering
	\includegraphics[width=1\linewidth]{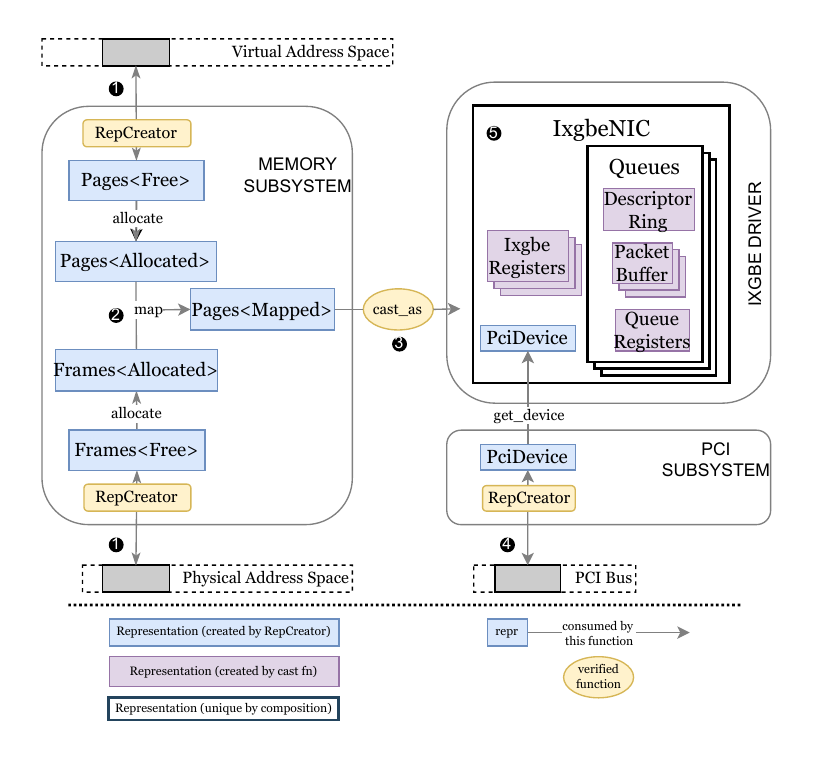}
 \vspace{-3em}
 \caption{\emph{The Interplay of an IRS and Formal Verification in \name:} We create representations for \code{Pages}, \code{Frames} and \code{PciDevice}s using the \code{\RepresentationCreator}, ensuring there is a 1-to-1 mapping between the hardware resource and the software representation \whitecircle{1}, \whitecircle{4}. Functions consume instances of \code{Pages} and \code{Frames} to transition them into the next state and to maintain the 1-to-1 mapping \whitecircle{2}. Once PTEs are added, a \code{Pages} instance is in the \code{Mapped} state and its represented memory can be cast to other representations \whitecircle{3}. We formally verify the cast functions to uphold uniqueness. \code{IxgbeNIC} consists of multiple representations and is unique by Rule of Composition \whitecircle{5}.}
 \label{fig:representation_system}
 \vspace{-2mm}
\end{figure}

\subsection{Overview}

In this section we detail our implementation of the \code{\RepresentationCreator}, our changes to the memory subsystem of \name, and our 10 GbE \program{ixgbe} driver.
The \code{\RepresentationCreator} is a \code{struct} with verified methods to create representations. 
It maintains the 1-to-1 mapping between the physical resource and its software representation(s), shown as step {\small \raisebox{0.1em}{\textcircled{\footnotesize \bfseries \raisebox{-0.7pt}{1}}}} in \autoref{fig:representation_system}.
It is a prominent example of how we use formal verification to increase the \cert of an \ICS invariant.
By following the \emph{Rule of Reuse}, we not only reduce the lines of code for creating representations but also the lines of specification and proof. We implemented the \code{\RepresentationCreator} in 392 SLOC and its proof-to-implementation ratio is 1:56.

We used the \code{\RepresentationCreator} to instantiate \code{Pages} and \code{Frames} in the memory subsystem. \code{Pages} and \code{Frames} can be in different typestates; the \code{Mapped} state gives access to the represented memory ({\small \raisebox{0.1em}{\textcircled{\footnotesize \bfseries \raisebox{-0.7pt}{2}}}} in \autoref{fig:representation_system}). The map state transition function adds PTEs using an intralingual HAL.
By applying our hybrid approach to the memory management subsystem of \name, we provide a stronger guarantee of the bijective mapping between pages and frames, a core invariant of \name \cite{boos2020osdi}. 
The implementation effort for the memory subsystem changes was 1.3k SLOC with a proof-to-implementation ratio of 1:171. 
 
We implemented the \program{ixgbe} NIC driver of Theseus using our approach. The driver supports the Intel 10 GbE 82599 NIC.
Representations in the network driver are created in three ways: through the \code{\RepresentationCreator}, e.g. , \code{PciDevice},  through verified memory cast functions, e.g., \code{IxgbeRegisters}, or 
by applying the Rule of Composition, e.g. \code{IxgbeNIC} ({\whitecircle{3}},{\whitecircle{4}},{\whitecircle{5}} 
in \autoref{fig:representation_system}). 
The driver implementation was 2k SLOC, of which only 141 lines required verification. The proof-to-implementation ratio of the verified portion of the driver was 1:8.3. The complete breakdown along with the additional effort to write external specifications is given in \autoref{table:verification}.

We choose the NIC driver for three reasons. First, there is a rich literature about formally verifying its properties~\cite{pirelli2020tinynf-osdi, hawblitzel2014ironclad}. We will be able to tease out the strengths and weaknesses of our approach in providing similar guarantees. 
Second, the performance of the NIC driver is highly sensitive to overhead and can be easily quantified, which allows us to evaluate any overhead from the intralingual design. 
Third, after reviewing the bugs in the DPDK \program{ixgbe} driver, we found that a majority could be prevented by techniques presented in this paper (\autoref{table:dpdk_ixgbe_bugs}), motivating us to focus on a network driver to evaluate our intralingual design techniques.

\begin{table}[t]
\caption{We found that approximately 50\% of bugs in the DPDK \program{ixgbe} driver ~\cite{dpdk_ixgbe_bugs} can be eliminated with our hybrid approach. We classified bugs based on the correctness technique that can be used to prevent them. 
A complete list of the bugs with the proposed solution for each can be found in \autoref{app:dpdk_ixgbe_bugs}.
}
\vspace{-1em}
\footnotesize
\begin{tabular}{l|c}
\textbf{Prevention Technique}& \multicolumn{1}{l}{\textbf{Number of Bugs}} \\ \hline
Basic Rust          & 4 (13.8\%)                                  \\
Intralingual HAL    & 5 (17.2\%)                                  \\
IRS                 & 6 (20.7\%)                                  \\
Formal Verification & 12 (41.4\%)                                 \\
Hardware Issue      & 2 (6.9\%)                                   \\\hline
\end{tabular}
\label{table:dpdk_ixgbe_bugs}
\end{table}

\label{sec:impl-repr-creator}

\begin{figure}[t]
    \begin{center}
    \begin{lstlisting}%
        [language=Rust, 
        xleftmargin=.03\textwidth,
        belowcaptionskip=-5.5mm,
        % float, 
        %frame=single,
        caption={A \code{\RepresentationCreator} object provides a verified interface to create unique \capis. The postconditions of its public method specify that, if successful, the \capi did not overlap with any pre-existing \capi and its identifying information was added to the bookkeeping. The Prusti keyword \code{ensures} starts a post-condition, \code{old} returns the value of an argument at the beginning of a function, and \code{result} refers to the return value},
        label={lst:rep_creator}]%
    %
    
    pub struct RepCreator<T: ResourceIdentifier, R> { $\label{line:rep_creator}$
        // for reps created before heap initialization
        array: Option<StaticArray<T>>,
        list: List<T>,
        constructor: fn(&T) -> R $\label{line:rep_constructor}$
    }

    impl<T: ResourceIdentifier, R> RepCreator<T,R> {
        #[ensures(result.is_ok() ==> {
            forall(|i: usize| i < old(self.list.len()) ==> 
                !old(self.list.lookup(i)).overlaps(&id))
            &&
            result.is_ok() ==> self.list.lookup(0)) == id
        })]  
        pub fn create_unique_representation(&mut self, id: T) $\label{line:create_unique_rep}$
            -> Result<R, RepresentationCreationError> 
        {    
            if !self.list.elem_overlaps(id) {
                self.list.push(&id);
                Ok(self.constructor(&id))
            } ...    
        }
    }

    pub trait ResourceIdentifier: Copy { $\label{line:resource_identifier}$
        #[pure]
        fn overlaps(&self, other: &Self) -> bool;
    }\end{lstlisting}%
    
    \end{center}
\end{figure}

\subsection{Intralingual Representation System (IRS)}
An IRS uses a linear type system to shift resource management tasks to the compiler, but the guarantees it provides are weak if we don't prove the uniqueness of every representation. 
To simplify this task, we implemented a generic representation creator; a portion of the code is given in \autoref{lst:rep_creator}. The \capi creator is a generic \code{struct} (\autoref{line:rep_creator}) composed of verified bookkeeping data structures and a private constructor that creates a \capi when given an identifier (\autoref{line:rep_constructor}).
The constructor can only be called through a verified method of \code{\RepresentationCreator} in which we search the list of existing \capis to make sure an overlapping one has not already been instantiated (\autoref{line:create_unique_rep}).

The \code{\RepresentationCreator} uses two generic types: \code{R} is the type of the \capi, and \code{T} is the type of the identifier;  it contains all the information required to create a \capi but itself does not give access to any resource.
The former is a linear type, and the latter is a clonable type which implements the \code{ResourceIdentifier} trait (\autoref{line:resource_identifier}).
We created the \code{ResourceIdentifier} trait so that the system developer can define what it means for \capis of the same type to overlap. The \code{overlap} trait method is considered part of the specification, and the correctness of our invariant depends on its correct definition.

For example, a representation of a PCI device is a \code{PCIDevice} object and its identifier is its bus, device, and function (slot) numbers, collectively given by the type \code{PCILocation}. The implementation of \code{ResourceIdentifier} for \code{PCILocation} defines an overlap as when the bus, device and function numbers are equal. 
Before scanning the PCI bus, we instantiate a \code{\RepresentationCreator<PCILocation, PCIDevice>} object, and for every connected device we create a \capi of it through the \code{create\_unique\_representation} method.

\begin{table*}[t]
\caption{Code size and verification times for the formally-verified portions we contributed to Theseus. For the PCI crate and \program{ixgbe} driver which contain a large amount of unverified code, we only included the verified  functions as part of the implementation size.
}
\label{table:verification}
\footnotesize
\begin{tabular}{l|c|ccc|ccc}
\hline
                       & \textbf{Intralingual Spec} & \textbf{Prusti Spec} & \textbf{Prusti Proof} & \textbf{Impl} & \textbf{Rust Compilation} & \textbf{Prusti Processing} & \textbf{Verification} \\
                       & \textit{(SLOC)}            &                      & \textit{(SLOC)}       &               &                           & \textit{(s)}            &                       \\ \hline
Rust External Spec     & 0                          & 295& 0                     & 0             &                           -&                         -&                       -\\
Theseus External Spec  & 0                          & 14& 0                     & 0             &                           -&                         -&                       -\\
Representation Creator & 0                          & 66& 7                     & 392&                           9.05&                         18.98&                       190.74\\
Frame Allocator        & 13                         & 28& 0                     & 151           &                           9.60&                         9.04&                       69.65\\
Page Allocator         & 13                         & 24                   & 0                     & 142           &                           9.57&                         8.70&                       65.43\\
Memory Functions       & 15                         & 164& 8& 1076&                           41.33&                         71.29&                       459.08\\
PCI Functions          & 3                          & 2& 0                     & 36&                           10.68&                         5.94&                       21.83\\
\textsf{ixgbe} Driver           & 16                         & 41& 17                    & 141&                           16.03&                         11.91&                       114.54\\ \hline
\textbf{Total}         & \textbf{60}                & \textbf{634} & \textbf{32} & \textbf{1938} & \textbf{96.26}& \textbf{125.86}& \textbf{922.27} \\ \hline \end{tabular}\\ 
\end{table*}

\subsubsection{Proof Sketch: A Representation is Unique}

The uniqueness invariant of an IRS states: 
\emph{Every representation is unique: there is no overlap between representations of the same type}.

We use our hybrid approach for correctness (\autoref{sec:hybrid}) to present a prose proof of the uniqueness invariant, wherein we explicitly state where each proof technique is used.
We list the \lems required to prove the invariant, and next to each \lem we list the techniques used to prove it: formal verification (\pt{F}), the type system (\pt{T}) or intralingual spec assertion (\pt{IS}). 
We use a prose proof (\pt{P}) to tie multiple techniques together.
Our proof is based on the following three \lems.

\vspace{1ex}\noindent\textbf{Lemma 1.} A representation cannot be cloned or copied as it is a linear type and does not implement the \code{Clone} trait. \pt{[T, IS, P]}

\noindent\textbf{Lemma 2.} Functions that create a representation or mutate its resource identifier always prevent overlaps. \pt{[F]}

\noindent\textbf{Lemma 3.} The resource identifier field of a representation is never changed in unverified functions. \pt{[T, IS, P]}

These conditions prove that a representation is unique at the time of instantiation (by using the \code{\RepresentationCreator}). Then, for its lifetime, it cannot be duplicated or mutated in a way that would jeopardize its uniqueness.

\subsection{Memory Management}
\label{sec:impl-mem}

We use our hybrid approach to uphold the bijective mapping invariant of \name. Isolation in Theseus without the use of hardware address spaces relies upon this invariant always being upheld.
The invariant can be equivalently restated as \emph{a frame can only be present in a single PTE}.
The memory subsystem of \name only adds a PTE when creating a \code{Pages<Mapped>} instance, and only removes it when dropping the same instance. 
Thus, proving the bijective mapping invariant necessitates proving the correct construction and destruction of a \code{Pages<Mapped>} instance.

\subsubsection{Intralingual Design of the Memory Subsystem}
The functions to walk the page table and update PTEs are categorized as part of the intralingual HAL of the memory subsystem; a PTE can only be manipulated through a type-safe interface.
We use an \ICS to create a \code{Frames} and \code{Pages} type, instances of which are the unique \capi of a region of physical or virtual memory, respectively.
With typestate programming, we create four possible states for instances of the \code{Frames} and \code{Pages} types: \code{Free}, \code{Allocated}, \code{Mapped}, and \code{Unmapped}. 
\code{Pages<Mapped>} instances (which represent accessible memory) are used according to the rules of the Rust type system: one mutable reference or multiple immutable references, which corresponds to a single writer or multiple readers to the underlying memory, but not both at the same time.
A detailed code example of the \code{Pages} type is given in \autoref{lst:intralingual_design}.

The memory subsystem first creates and stores \code{Frames<Free>} and \code{Pages<Free>} instances during its initialization routine, once we have information about the size of the physical address space. 
A task will allocate \code{Frames} and \code{Pages} when it needs more memory; the allocate function transitions these instances to the \code{Allocated} state and returns them to the caller.
The mapping function consumes instances in an \code{Allocated} state, converts them to a \code{Mapped} state, and adds PTEs that associate each page represented by the \code{Pages} instance with one frame represented by the \code{Frames} instance in order to create an exclusive, bijective mapping. 
The mapping function returns the \code{Pages<Mapped>} instance to the caller such that it can be used to access the underlying memory, but forgets the \code{Frames<Mapped>} instance immediately as an efficiency optimization. 
The pages represented by a \code{Pages<Mapped>} instance are unmapped upon being dropped,
during which the drop handler clears the associated PTEs and transitions it to the \code{Unmapped} state. The drop handler also recreates the previously-forgotten \code{Frames<Unmapped>} instance from information stored in the PTEs.
Once their TLB entries are invalidated, the \code{Frames} and \code{Pages} are transitioned to the \code{Allocated} state and then dropped, 
which subsequently transitions them to the \code{Free} state, which finally returns them to the allocator to be stored for future use.

\subsubsection{Hybrid Approach to Uphold Uniqueness}
\label{sec:impl-mem-hybrid}
We use formal verification to uphold the uniqueness guarantee of \code{Frames} and \code{Pages}, and to prove that functions which cast a pointer lying in the page range of a \code{Pages<Mapped>} also uphold uniqueness.

For \code{Pages} and \code{Frames}, we introduce the linear type \code{Chunk} that is initially created through the \code{\RepresentationCreator}.
The resource identifier is a \code{RangeInclusive<usize>}.
We also formally verify the methods \code{split()} and \code{merge()} that create \code{Chunk} instance(s) by consuming existing ones. 
We make its inner field private to prevent mutable access outside of its methods, and use static assertions to make sure a \code{Chunk} instance is only mutably accessible from its verified methods.
These joint properties of \code{Chunk} prove that every instance is unique.
Both \code{Frames} and \code{Pages} are composed of only the \code{Chunk} type, for which the static assertion:\\
\centerline{ \small \code{assert\_fields\_type!(Pages: range: Chunk);}}\\
ensures this composition relationship always holds (\autoref{sec:hybrid-intralingual-spec}). Consequently, \code{Frames} and \code{Pages} are unique by the Rule of Composition, ensured by the Rust language --- an example of how the type system can extend formally-verified invariants to other types (\autoref{sec:hybrid-rules}).

We verify the cast functions to prove that the returned reference is unique.
The cast functions consume a \code{Pages<Mapped>}, so its uniqueness invariant serves as a pre-condition to the functions.
To prove uniqueness of the newly created representation, we only need to prove that the returned reference (and the memory we access through it) lies within the given page range. If there are multiple pointer casts from within the same page range, we prove that the instances they point to do not overlap.
Since we cannot verify unsafe code, we separate a casting function into two. 
The first is verified and returns the address of the pointer. The second function is trusted and consists of a single line of unsafe code that actually performs the cast.
The cast functions are generic and can be used to cast untyped memory to an instance of any ``plain old data'' type following the Rule of Reuse.

\subsubsection{Proof of Uniqueness: a Bug Revealed}
\label{sec:bug}
The proof of uniqueness of \code{Pages} and \code{Frames} greatly increases the \cert of the bijective mapping invariant.
The previous reasoning behind this invariant was that the \code{map} function consumes both a \code{Pages<Allocated>} and \code{Frames<Allocated>} instance, assuming both are unique. It then adds PTEs for them and returns a \code{Pages<Mapped>} instance. 
This uniqueness property is no longer just an assumption, instead a valid pre-condition to the \code{map} function. 

The original \name relied on manual checks in place of formal verification to prove this invariant, as uniqueness is beyond the scope of intralingual design. 
In the frame allocator bookkeeping code, we discovered a bug that led to the instantiation of {\em overlapping} \code{Frames} instances, violating the bijective mapping invariant.
This bug only manifested in a very particular code path that had not occurred in over four years of frame allocator code usage.
This bug stalled network driver development for over one person-month,
motivating us to explore ways to incorporate formal verification into an intralingual system. 

\subsection{\textsf{ixgbe} Driver}
\label{sec:implementation-driver}

We implemented the ixgbe driver for the Intel 82599 NIC. Our goal was to write a "correct by construction" driver where most bugs could be caught at compile time. We define correctness properties, then detail how intralingual design and formal verification work together to uphold them.
Through a review of previous works \cite{hawblitzel2014ironclad, pirelli2020tinynf-osdi}, we can broadly classify correctness into three properties.

\textbf{P.1.} \emph{Datasheet-compliant bit-level communication:} When accessing the MMIO registers or data structures, the driver must enforce the read/write restrictions for every field. Past work ensures this property either through IDLs \cite{merillon2000osdi, conway2004ndl, sun2005hail} or symbolic execution \cite{pirelli2020tinynf-osdi}.

\textbf{P.2.} \emph{Creation and management of software resources:} The driver uses data structures, e.g., a ring buffer of packet descriptors, to communicate with the device, and it must map them with the required alignment and length. The underlying memory must be accessible to the driver and device, and be returned to the OS when no longer in use.

\textbf{P.3.} \emph{Valid bookkeeping state:} The driver maintains bookkeeping state such as the index of the next descriptor to use, the filter table that stores forwarding rules, etc.
The driver provides an interface to the OS to use device functions, and after every interface call, the driver must update its bookkeeping state to accurately reflect the device state.

 We use an intralingual HAL to uphold \textbf{\small P.1.} and use IRS and formal verification to uphold \textbf{\small P.2.} and \textbf{\small P.3.}.

\subsubsection{Intralingual Design of the \program{ixgbe} Driver}

The ixgbe driver uses an intralingual HAL to communicate with the device. It prevents bugs at the driver{$\leftrightarrow$}device interface at compile time \textbf{\small{(P.1.)}}, as shown (in part) in \autoref{lst:intralingual_hal}.

We implement an IRS in the driver that shifts some resource management to the compiler, helping to maintain \textbf{\small{P.2.}}. 
In the initialization code, the driver 
creates a software representation of each physical NIC (\code{IxgbeNIC}), which is composed of representations of receive queues (\code{RxQueue}), transmit queues (\code{TxQueue}), and device registers.
Receive and transmit queues are further composed of descriptor rings, packet buffers, and queue registers. This compositional relationship is shown in the ixgbe driver module of \autoref{fig:representation_system}.
The driver creates descriptor rings, packet buffers and registers through a verified interface, so \code{RxQueue} and \code{TxQueue} representations are unique by the \emph{Rule of Composition}.

We create representations to descriptor rings, packet buffers, and queue registers by using a cast function which takes ownership of a \code{Pages<Mapped>} instance and casts an untyped pointer within the page range to a Rust reference with the type of the representation; the lifetime of the reference is tied to the \code{Pages<Mapped>} instance. In \autoref{fig:representation_system}, the \code{cast\_as} function consumes \code{Pages<Mapped>} instances to create  certain representations (marked in purple).
The driver's ownership of the \code{Pages<Mapped>} instance inherently proves that it has access to the represented memory \textbf{\small{(P.2.)}}.
When representations created through a cast function are dropped, the \code{Pages<Mapped>} instances they are composed of are also dropped. 
The drop handler unmaps the pages and returns them to the OS \textbf{\small{(P.2.)}}.

\code{IxgbeNIC} also owns its \code{PCIDevice} representation, so that no other entity in the system can access its PCI configuration space once it is initialized.
The drop handler for \code{PCIDevice} returns the representation to the OS so that the NIC can be initialized again.
With this design, the \code{IxgbeNIC} representation is unique by the Rule of Composition and its methods are the only way to communicate with the device. 
With the IRS, we can now rely on Rust's inbuilt rules to prevent any race conditions or illegal accesses to device memory.

We use typestates to prevent the OS from accessing disabled device features and writing meaningless values to bookkeeping state \textbf{\small{(P.3.)}}.
We implement typestates for the \code{RxQueue} and \code{TxQueue} types; both can be in either the \code{Enabled} or \code{Disabled} state.
In the \code{Disabled} state, the \code{send()} and \code{receive()} methods are not exposed.
In addition, the \code{RxQueue} has two additional states: \code{L3/L4 Filter} and \code{RSS}, representing two different NIC features: 5-tuple filters and Receive Side Scaling (RSS), which are used to distribute incoming packets among receive queues.
These features are important enough to merit their own typestates because as NIC bandwidth increases, it can only be fully utilized by employing multiple queues.
Different filter types are also mutually exclusive, and a queue used for RSS must not be used by a 5-tuple filter \cite{82599datasheet}. These misconfiguration errors commonly lead to confusion about the destination queue of a packet \cite{dpdk399}, and our use of typestates prevents them.

\subsubsection{Hybrid Approach to Correctness Properties}

We use formal verification to help uphold \textbf{\small P.2.} and \textbf{\small P.3.}.
The driver creates multiple representations using verified functions that cast untyped memory of a \code{Pages<Mapped>} (\autoref{sec:impl-mem-hybrid}).
For the driver to completely uphold \textbf{\small P.2.}, we updated the verification to also prove alignment invariants in addition to uniqueness.
The alignment and uniqueness properties of the newly created representations are now \emph{type invariants}, and we never need to re-verify these properties following the Rule of Type Preservation. 

We also formally verify core functions that update bookkeeping state to uphold \textbf{\small P.3.}.
We verify the \code{send\_batch} and \code{receive\_batch} functions to prove that the value of the next descriptor to use always lies within the descriptor ring.
We verify the \code{add\_filter} function to prove that a new filter is added with the given IP addresses and port numbers, as long as there is an unused filter slot (the 82599 NIC only offers 128) and no existing identical filter. It is the software driver's responsibility to prevent this logical error.
Since we limit verification to a few functions in the driver module, we need to make sure that other functions within the module do not break the verified invariants. For that, we add the \code{\#[nomutates]} macro at the top of unverified functions to ensure they cannot mutate bookkeeping state.

\section{Evaluation}
\begin{figure}[t]
	\centering
	\includegraphics[width=0.9\linewidth,trim={0.2cm 0.7cm 0.2cm 0.7cm},clip]{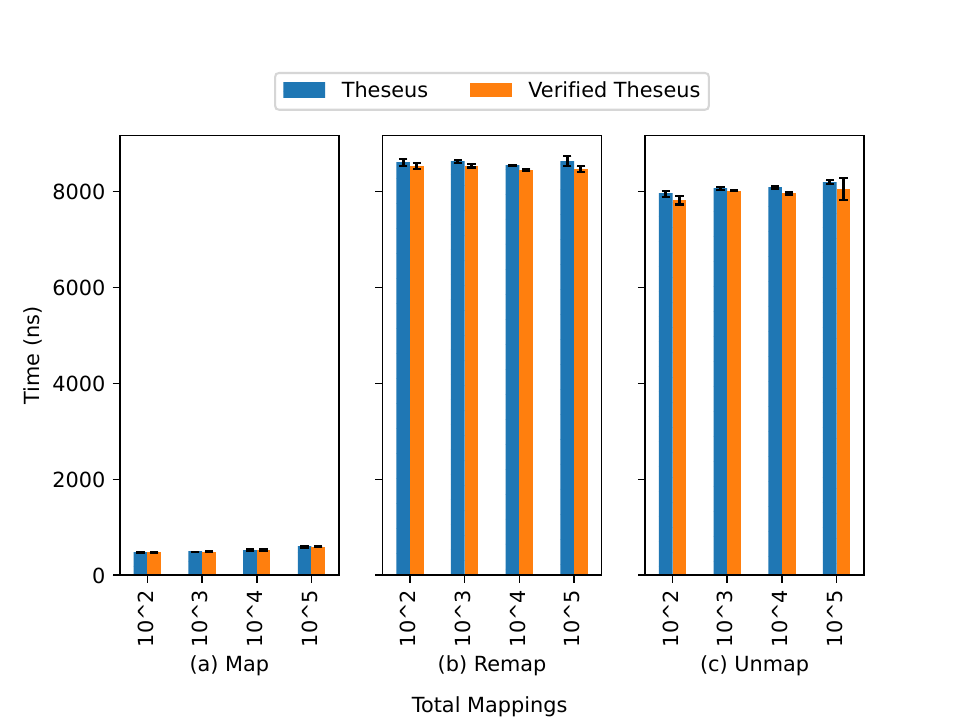}
 \vspace{-2mm}
	\caption{The individual time to map, remap, and unmap a 4 KiB page does not increase when verification is added to \name. The results presented are the mean times for 1 page, with the error bars representing the standard deviation.}
 \label{fig:memory}
 \vspace{-2mm}
\end{figure}
\begin{figure}[t]
	\centering
	\includegraphics[width=0.95\linewidth,trim={1cm 0 2cm 0.6cm},clip]{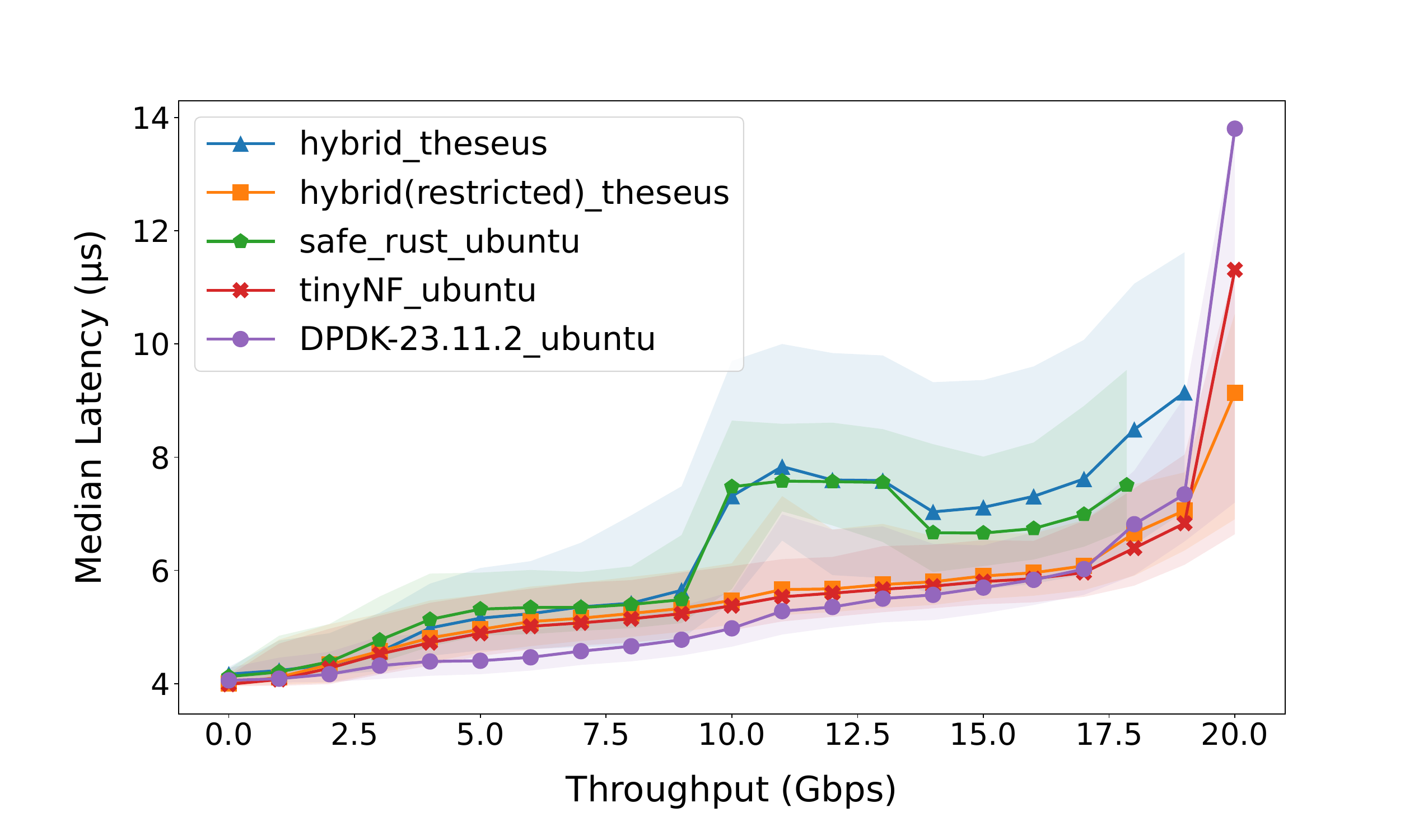}
 \vspace{-2mm}
]	\caption{A performance comparison of \program{ixgbe} drivers demonstrates that the hybrid driver has only 5\% lower maximum throughput than DPDK and has a latency profile similar to other research drivers like safe\_rust~\cite{pirelli2023safe}. The observed latency spike at 10 Gbps is likely a hardware issue and has been reported in previous works~\cite{pirelli2020tinynf-osdi,pirelli2023safe}. In the legend, the OS for each driver is indicated alongside its name. The shaded region illustrates the range between the 5th and 95th percentiles.}
 
 \label{fig:driver_latencies}
 \vspace{-2mm}
\end{figure}

\label{sec:evaluation}
We evaluate the performance of the updated memory subsystem of \name and the \program{ixgbe} network driver, and quantify the verification effort. 
Our objective is to find any performance difference caused by code changes resulting from the verification process. We assess the proof effort needed to formally verify sections of the memory subsystem and the network driver to gauge the ``lightweightedness'' of the hybrid approach, partly by comparing our network driver with formally verified network drivers reported in the literature, namely Ironclad~\cite{hawblitzel2014ironclad} and TinyNF~\cite{pirelli2020tinynf-osdi}. Additionally, we test the effectiveness of an intralingual HAL by inserting it as the hardware interface layer for other Rust \program{ixgbe} drivers. 

\subsection{Performance Comparison}

\paragraph{Setup} All our measurements were collected on an Intel(R) Xeon(R) Gold 6252N CPU at 2.30 GHz with hyper-threading disabled. To measure the Ethernet driver performance,  
we employ the RFC2544 benchmarking setup, which involves two machines (a device-under-test and a tester) connected by two 10 Gbps links. Each machine has two Intel 82599 Ethernet Controllers, with one port per NIC in use. On the device-under-test, we run the forwarder on either Theseus (to test our driver) or Ubuntu 18.04 (for the comparison driver). The tester machine runs the MoonGen Packet Generator in a VM with PCI passthrough to the NIC. We first run the RFC2544 zero packet loss test, used by DPDK~\cite{dpdkperf}, to find the maximum bidirectional throughput the driver could handle. 
Then, in line with recent works on driver design~\cite{pirelli2020tinynf-osdi,pirelli2023safe}, we measure round-trip latency as background traffic increases from 0 to the maximum throughput in 1 Gbps increments.

\paragraph{Memory Subsystem} We find that \name with formally-verified code (\textit{Verified \name}) performs similarly to the original \name that had no verified code.
We run two memory subsystem microbenchmarks. The first microbenchmark is a Rust version of  LMBench's~\cite{mcvoy1996lmbench} memory map. In this benchmark, a 4~KiB page is mapped, written, and unmapped 100,000 times. Both versions of \name show identical performance, a mean time of 1.99~$\mu$s with a standard deviation less than the timer period (42 ns).
The second microbenchmark is taken from the original \name paper~\cite{boos2020osdi}, which separately measures the time to map a page, remap it, and then unmap it, with an increasing number of mappings. \autoref{fig:memory} shows no significant difference between the two versions.

\paragraph{\program{ixgbe Driver}}
We find that the performance of \program{ixgbe} drivers written with our hybrid approach is comparable to that of the DPDK ixgbe driver, which does not come with any correctness guarantee.
We test two versions of our driver: \program{ixgbe\_hybrid}, a standard driver where packet buffers can be used in any order after receipt, and \program{ixgbe\_hybrid(restricted)}, which sends packet buffers in the order they are received. 
We implement the restricted version because TinyNF~\cite{pirelli2020tinynf-osdi} demonstrated that a driver using the restricted model is simple enough to be amenable to verification and can also achieve maximum throughput. 
To have a fair comparison we need to use the same driver model.

In ~\autoref{fig:driver_latencies} we show the packet round-trip latency as traffic throughput increases. The last latency measurement is taken at the maximum throughput that can be handled by the driver. \program{ixgbe\_hybrid} achieves a maximum bidirectional throughput of 19 Gbps, only 5\% lower than the 20 Gbps achieved by the DPDK ixgbe driver. DPDK is a highly optimized driver that uses SIMD instructions to process packets in batches and can maintain a lower round-trip latency per-packet than \program{ixgbe\_hybrid}. 
\program{ixgbe\_hybrid(restricted)} can handle the NIC's maximum throughput of 20Gbps due to fewer operations per packet.
\program{ixgbe\_hybrid(restricted)}, the TinyNF ixgbe driver, and the DPDK ixgbe driver all have similar latency profiles as was reported previously for the restricted driver model by TinyNF~\cite{pirelli2020tinynf-osdi}.

\begin{table*}[!t]
\caption{Bugs found in the ixgbe drivers from three Rust-based OSes that can be prevented by the intralingual HAL. For some bugs, identical versions were found in the DPDK driver using symbolic execution \cite{zaostrovnykh2019sosp}.}
\vspace{-0.5em}
\footnotesize
\resizebox{\textwidth}{!}{%
\begin{tabular}{l|ccc|c|c}
\hline
\multicolumn{1}{c|}{\multirow{2}{*}{\textbf{Bug Description}}}                    & \multicolumn{3}{c|}{\textbf{Present in Driver}}                                                                             & \multicolumn{1}{c|}{\multirow{2}{*}{\textbf{\begin{tabular}[c]{@{}c@{}}Intralingual HAL \\ Solution\end{tabular}}}} & \multirow{2}{*}{\textbf{DPDK ID}} \\
\multicolumn{1}{c|}{}                                                             & \multicolumn{1}{c}{\textbf{\scriptsize{ixy~\cite{emmerich2018ixy}}}}          & \multicolumn{1}{c}{\textbf{\scriptsize{Redox~\cite{redox}}}}            & \textbf{\scriptsize{Redleaf~\cite{narayanan2020redleaf}}}              & \multicolumn{1}{c|}{}                                                                                               &                                    \\ \hline
Write to reserved bit of EIMC register                                             & \multicolumn{1}{c}{\checkmark} & \multicolumn{1}{c}{}                          & \checkmark & register access function                                                                                            & 23                                 \\ 
Write to reserved bits of DTXMXSZRQ register                                       & \multicolumn{1}{c}{\checkmark} & \multicolumn{1}{c}{\checkmark} & \checkmark & register access function                                                                                            &                                    \\ 
RDRXCTL register is not set to default value                                       & \multicolumn{1}{c}{\checkmark} & \multicolumn{1}{c}{\checkmark} & \checkmark & register access function                                                                                            & \multicolumn{1}{l}{}              \\ 
Write to FCTRL register without clearing RXCTRL.RXEN                      & \multicolumn{1}{c}{\checkmark} & \multicolumn{1}{c}{\checkmark} & \checkmark & linear type as a proof of work                                                                                      & 21                                 \\
\hline
\end{tabular}
}
\label{tab:ixgbe-hal-bugs}
\end{table*}

\subsection{Verification and Implementation Effort}
\autoref{table:verification} reports the size (in SLOC) and verification times for the verified additions to \name. 
We find that the proof effort is magnitudes lower than end-to-end formally verified systems, and verification times are within minutes. 
All verification times are measured using the Prusti 2023-08-22 release. We time the verification of each crate individually, only exporting specifications from dependencies without verifying them. To identify where Prusti spends most of its time, we separately measure the time taken by the Rust compiler, the generation of verification conditions by Prusti, and the runtime for the Viper verification back end.

The formally-verified portion of \name consists of:
\emph{(i)} external specifications for types and functions from the Rust core library and for select crates in \name,
\emph{(ii)} a generic representation constructor including verified data structure implementations,
\emph{(iii)} portions of the page and frame allocator code that include methods to create and modify a \code{Chunk}, 
\emph{(iv)} definitions of memory related \code{struct}s and memory functions that take a \code{Pages<Mapped>} and cast a pointer in its page range,
\emph{(v)} select functions in the PCI crate,
and \emph{(vi)} select \program{ixgbe} driver functions.

We next compare the proof effort required for our lightweight correctness guarantees to that for stronger correctness guarantees using end-to-end verification, e.g., Ironclad~\cite{hawblitzel2014ironclad},  and  symbolic execution, e.g., TinyNF~\cite{pirelli2020tinynf-osdi}.

\paragraph{Proof Effort in the Memory Subsystem Compared}
The proof-to-implementation ratio for changes to the memory subsystem of \name (including the representation creator code) is \code{1:117}.
This is lower than the \code{10:1} proof-to-implementation ratio of an end-to-end verified page table implementation written in Rust \cite{brun2023beyond}, by more than two orders of magnitude. 
This minimal proof effort is a direct result of our hybrid approach,
and comes at the cost of lower \cert and a larger TCB.
We only use SMT verification for one property (uniqueness), and use other techniques to reason about the correctness of writes to the page table.

\paragraph{Proof Effort for the \program{ixgbe} Driver Compared}
The verified part of the hybrid \program{ixgbe} driver has a proof-to-implementation ratio of 1:8.3, less than the 1:4.8 ratio of the Ironclad apps Intel 82541PI 1 GbE driver~\cite{hawblitzel2014ironclad}, the only end-to-end verified Ethernet driver we could find. 
The 82541PI is a simpler device but the driver structure is similar,
and we make sure to prove the same invariants for the hybrid driver.
A strength of the hybrid approach is that it significantly reduces the lines of specification by verifying type invariants once and relying on the type system to carry them forward. 
Our hybrid method achieves a specification-to-implementation ratio of 1:3.4, compared to 1:2 for the Ironclad driver. 
Unlike the Ironclad driver, the hybrid driver does not add specification to check invariants for the \code{PciDevice}, \code{RxQueue}, \code{TxQueue} types since they have already been proven in the type constructors.

\paragraph{Intralingual HAL vs a Hardware Model}
We compare the effort that goes into writing an intralingual HAL with that of writing a hardware model and show that they take equivalent effort to encode the same subset of device behavior.
The TinyNF \program{ixgbe} driver itself does not contain any verified functions. Instead, invariants about it are proven by symbolically executing it against a 82599 hardware model. The hardware model contains assertions about bit-level communication with the device. Vigor used the same hardware model to find bugs in the DPDK ixgbe driver through symbolic execution~\cite{zaostrovnykh2019sosp}.

The \program{ixgbe\_hybrid(restricted)} driver's HAL is \emph{634 SLOC}, while the hardware model is 1.2k SLOC, though the register map and relevant functions for TinyNF are only \emph{615 SLOC}. 
With the intralingual HAL, we only need to encode the data sheet information once, within the implementation. Hardware models require double the effort, encoding the same information in both the model and the driver.

\paragraph{Overhead of Intralingual Spec}
We added just 60 lines of intralingual specification to prevent code changes from compromising the type-based properties that uphold invariants for the memory subsystem and the \program{ixgbe} driver. The specification includes preventing the implementation of \code{Clone} and \code{DerefMut} for base representation types and maintaining the compositional relationship between representations. Any \code{struct} field updated through formally verified methods is made private, typestate transition methods always consume representations in their previous state, and unverified functions that take mutable references are prevented from modifying verified values. It is easier to review these 60 lines of assertions than to search for the properties in the thousands of lines of code in the memory and \program{ixgbe} crates, making our hybrid approach more maintainable.

\subsection{Bugs Prevented}
Most techniques from \autoref{sec:intralingual} and \autoref{sec:hybrid} help write ``correct by construction'' code to prevents bugs at implementation time. Nevertheless, the intralingual HAL is portable and can be used to find bugs by retrofitting it into an existing implementation. We inserted the \program{ixgbe} intralingual HAL as the driver-hardware interface in the \program{ixgbe} drivers from three other open-source Rust OSes, namely, Redleaf~\cite{narayanan2020redleaf}, Redox~\cite{redox}, and ixy~\cite{emmerich2018ixy}. It revealed four previously unreported bugs as detailed in \autoref{tab:ixgbe-hal-bugs}. Three were present in all the drivers most likely because the Redleaf and Redox drivers are adapted from ixy. We submitted fixes for these bugs, which have already been accepted into the corresponding mainlines. 

\section{Concluding Remarks}
\label{sec:conclusion}
Using the Theseus operating system as an experimental ground, we show that Rust's type system can be used to provide weaker but useful correctness guarantees for system software, especially when combined with formal verification in a hybrid approach. The key is to cleverly exploit the intralingual design patterns described in \autoref{sec:intralingual} to expand the reach of guarantees from formally-verified code, by using rules and intralingual specifications described in \autoref{sec:hybrid}. Our approach makes previously impossible trade-offs between proof effort, \cert, and size of trust base. In practice, it has already prevented and discovered bugs in multiple Rust-based operating systems. We hope that sharing our experience with the community will help develop it further and present/discover bugs in the growing body of Rust software systems. 

\section*{Acknowledgments}
This work is supported in part by National Science Foundation Award \# 2416594.

\bibliographystyle{plain}
\bibliography{abr-short,paper-specific,driver,embedded,os,formal,history,reference-count,zhong,dpdk-bugs,rust}
\newpage
\appendix
\section{Classification of DPDK Ixgbe Driver Bugs}
\label{app:dpdk_ixgbe_bugs}

In \autoref{tab:dpdk_ixgbe_bugs_desc} we detail DPDK ixgbe bugs \cite{dpdk_ixgbe_bugs} and the technique that can be used to prevent them. 
At the time of compiling this list, there were 93 reported bugs which referenced the ixgbe driver.
Of those 93, 29 were actually related to the driver. 
The remaining bugs were either invalid, or concerned with other parts of the DPDK framework and not the driver specifically. A condensed version of this table is given in the paper.

\begin{table*}[t]
\resizebox{\textwidth}{!}{%
\begin{tabular}{|c|l|l|}
\hline
\textbf{Bug ID} & \multicolumn{1}{c|}{\textbf{Notes}}                                                                                                                                                                                                     & \multicolumn{1}{c|}{\textbf{Resolution}} \\ \hline
21              & "Ixgbe driver changes FCTRL without first disabling RXCTRL.RXEN"                                                                                                                                                                        & Intralingual HAL                         \\ \hline
22              & "Ixgbe driver sets RDRXCTL with the wrong RSCACKC and FCOE\_WRFIX values"                                                                                                                                                               & Intralingual HAL                         \\ \hline
23              & "Ixgbe driver writes to reserved bit in the EIMC register"                                                                                                                                                                              & Intralingual HAL                         \\ \hline
25              & "Ixgbe driver sets TDH register after TXDCTL.ENABLE is set"                                                                                                                                                                             & Intralingual HAL                         \\ \hline
26              & "Ixgbe driver does not ensure FWSM firmware mode is valid before using it"                                                                                                                                                              & Intralingual HAL                         \\ \hline
57              & Null pointer de-reference                                                                                                                                                                                                               & Basic Rust                               \\ \hline
69              & Maximum wait time for link to come up was too small                                                                                                                                                                                     & Formal verification                      \\ \hline
103             & Deadlock during initialization                                                                                                                                                                                                          & Formal verification                      \\ \hline
116             & \begin{tabular}[c]{@{}l@{}}Segmentation fault when in-use rx queues are freed. \\ This can be prevented by an IRS.\end{tabular}                                                                                                         & Intralingual representations             \\ \hline
216             & Burst size should be \textgreater{}= 4 to use a vectorized function                                                                                                                                                                     & Formal verification                      \\ \hline
263             & "ixgbe does not support 10GBASE-T copper SFP+"                                                                                                                                                                                          & Hardware issue                           \\ \hline
350             & "ixgbe: incorrect speed capabilities advertised for X553 devices"                                                                                                                                                                       & Formal verification                      \\ \hline
372             & Driver needs to separately handle different error cases.                                                                                                                                                                                & Formal verification                      \\ \hline
388             & "ixgbe: link state race condition can occur when starting a fiber port"                                                                                                                                                                 & Basic Rust                               \\ \hline
399             & \begin{tabular}[c]{@{}l@{}}Filtering and RSS are enabled at the same time, leading to confusing results. \\ Typestates would prevent enabling both features simultaneously.\end{tabular}                                                & Intralingual representations             \\ \hline
447             & Resource cleanup always occurs with representation drop handlers                                                                                                                                                                        & Intralingual representations             \\ \hline
514             & Runtime flag check does not check for IPv6 flag                                                                                                                                                                                         & Formal verification                      \\ \hline
516             & Vectorized functions should receive the specified number of packets, but always receive 32                                                                                                                                              & Formal verification                      \\ \hline
629             & Device marks checksum as invalid.                                                                                                                                                                                                       & Hardware issue                           \\ \hline
643             & \begin{tabular}[c]{@{}l@{}}Rx queue was initialized even though failed to allocate packet buffers. \\ With an IRS, it would be impossible to create an rx queue unless packet buffers \\ are created and then owned by it.\end{tabular} & Intralingual representations             \\ \hline
650             & Improvements to prevent packet loss                                                                                                                                                                                                     & Formal verification                      \\ \hline
664             & Reta table can be set before the device is started. Typestates would prevent this.                                                                                                                                                      & Intralingual representations             \\ \hline
869             & Use after free                                                                                                                                                                                                                          & Basic Rust                               \\ \hline
882             & Bug in receive function lets application use packet buffer that is still in descriptors                                                                                                                                                 & Formal verification                      \\ \hline
1034            & Enabled IPv4 checksum offload even though device doesn't support it.                                                                                                                                                                    & Formal verification                      \\ \hline
1057            & \begin{tabular}[c]{@{}l@{}}A single API to set flow rules for different drivers, leads to setting invalid flow rules \\ for some devices. Strongly typed interfaces which prevent invalid arguments would prevent this.\end{tabular}    & Basic Rust                               \\ \hline
1106            & Missing OR operator                                                                                                                                                                                                                     & Formal verification                      \\ \hline
1249            & Missing NEGATION operator                                                                                                                                                                                                               & Formal verification                      \\ \hline
1259            & \begin{tabular}[c]{@{}l@{}}Port is restarted without restarting queues. \\ Typestates for representations would prevent this.\end{tabular}                                                                                              & Intralingual representations             \\ \hline
\end{tabular}%
}
\caption{List of DPDK ixgbe bugs }
\label{tab:dpdk_ixgbe_bugs_desc}
\end{table*}

\end{document}